\documentclass[
aps,
pra,
showpacs,
amsmath,
amssymb,
preprint,
floatfix,
lengthcheck, 
superscriptaddress,
longbibliography
]{revtex4-1}

\usepackage{graphicx}
\usepackage{mathtools}
\usepackage{braket}
\usepackage{hyperref}
\usepackage{bbold}
\usepackage{calrsfs}
\usepackage{dsfont}
\usepackage{cancel}
\usepackage{mathrsfs}
\usepackage{multirow}
\usepackage{latexsym}
\usepackage[normalem]{ulem}
\usepackage{float}

\usepackage{color}

\DeclareMathAlphabet{\pazocal}{OMS}{zplm}{m}{n}

\begin{document}


\title{Frequency-dependent Sternheimer linear-response formalism for strongly coupled light-matter systems}

\author{Davis M. Welakuh}
\email[Electronic address:\;]{dwelakuh@seas.harvard.edu}
\affiliation{Max Planck Institute for the Structure and Dynamics of Matter and Center for Free-Electron Laser Science \& Department of Physics, Luruper Chaussee 149, 22761 Hamburg, Germany}
\affiliation{Harvard John A. Paulson School Of Engineering And Applied Sciences, Harvard University, Cambridge, Massachusetts 02138, USA}

\author{Johannes Flick}
\email[Electronic address:\;]{jflick@flatironinstitute.org}
\affiliation{Center for Computational Quantum Physics, Flatiron Institute, 162 5th Avenue, New York, NY 10010, USA}

\author{Michael Ruggenthaler}
\email[Electronic address:\;]{michael.ruggenthaler@mpsd.mpg.de}
\affiliation{Max Planck Institute for the Structure and Dynamics of Matter and Center for Free-Electron Laser Science \& Department of Physics, Luruper Chaussee 149, 22761 Hamburg, Germany}
\affiliation{The Hamburg Center for Ultrafast Imaging, Luruper Chaussee 149, 22761 Hamburg, Germany.}

\author{Heiko Appel}
\email[Electronic address:\;]{heiko.appel@mpsd.mpg.de}
\affiliation{Max Planck Institute for the Structure and Dynamics of Matter and Center for Free-Electron Laser Science \& Department of Physics, Luruper Chaussee 149, 22761 Hamburg, Germany}

\author{Angel Rubio}
\email[Electronic address:\;]{angel.rubio@mpsd.mpg.de}
\affiliation{Max Planck Institute for the Structure and Dynamics of Matter and Center for Free-Electron Laser Science \& Department of Physics, Luruper Chaussee 149, 22761 Hamburg, Germany}
\affiliation{Center for Computational Quantum Physics, Flatiron Institute, 162 5th Avenue, New York, NY 10010, USA}
\affiliation{The Hamburg Center for Ultrafast Imaging, Luruper Chaussee 149, 22761 Hamburg, Germany.}


\begin{abstract}
The rapid progress in quantum-optical experiments especially in the field of cavity quantum electrodynamics and nanoplasmonics, allows to substantially modify and control chemical and physical properties of atoms, molecules and solids by strongly coupling to the quantized field. Alongside such experimental advances has been the recent development of ab-initio approaches such as quantum electrodynamical density-functional theory (QEDFT) that is capable of describing these strongly coupled systems from first-principles. To investigate response properties of relatively large systems coupled to a wide range of photon modes, ab-initio methods that scale well with system size become relevant. In light of this, we extend the linear-response Sternheimer approach within the framework of QEDFT to efficiently compute excited-state properties of strongly coupled light-matter systems. Using this method, we capture features of strong light-matter coupling both in the dispersion and absorption properties of a molecular system strongly coupled to the modes of a cavity. We exemplify the efficiency of the Sternheimer approach by coupling the matter system to the continuum of an electromagnetic field. We observe changes in the spectral features of the coupled system as Lorentzian line shapes turn into Fano resonances when the molecule interacts strongly with the continuum of modes. This work provides an alternative approach for computing efficiently excited-state properties of large  molecular systems interacting with the quantized electromagnetic field.
\end{abstract}

\maketitle


\section{Introduction}

Strong interactions between light and matter within enhanced photonic environments such as optical cavities and plasmonic devices have attracted great attention in recent years. The signature of such strong interactions is the formation of new hybrid light-matter states (polaritons) which are manifested by a Rabi splitting in the spectrum of the coupled system. These new states of matter can be used to control, for instance, chemical reactions~\cite{hutchison2012,thomas2016,schaefer2021}, enhance charge and energy transport~\cite{coles2014,zhong2016,feist2015}, selectively manipulate electronic excited-states~\cite{stranius2018}, to name a few examples. Such coupled light-matter systems have the tendency to exhibit significantly different properties than the uncoupled subsystems even at ambient conditions, which suggests various interesting applications in chemistry and material science~\cite{chikkaraddy2016,bellessa2004,ebbesen2016,santhosh2016}. These intriguing effects caused by the emergence of polaritons manifest strongly in the excited-state properties of the coupled systems, for example, in the absorption or emission spectra~\cite{chikkaraddy2016,bellessa2004,ebbesen2016,stranius2018}. 

The occurrence of different effects due to the emergence of polaritons shows the complexity that arises when light and matter strongly mix. Due to this inherent complexity of the coupled light-matter system, the theoretical description of these effects are non-trivial. Quite often, the coupled system is studied with quantum optical models that potentially over-simplify the matter subsystem. One of such simplifications selects just a few energy-levels of an atomic or molecular system and couples it to the photon modes of an optical cavity~\cite{dicke1954,jaynes1963,garraway2011}. Another common simplification is on the photon side where a realistic cavity which is normally open is reduced to just a few modes that cannot account for the finite lifetime of excitations. However, in many cases these phenomenological models are not sufficient to capture important details of the coupled system, for instance, the emergence of bound states in the continuum~\cite{sidler2020}, how collective strong coupling leads to local modification of chemical properties~\cite{sidler2021} and in cavity-modified chemistry where the reaction-rate is reduced under cavity induced resonant vibrational strong coupling~\cite{thomas2016,schaefer2020}. This calls for ab initio methods which allow to treat from first-principles complex matter systems interacting with the quantized field~\cite{ruggenthaler2014,flick2015,haugland2020} within non-relativistic quantum electrodynamics (QED)~\cite{spohn2004,ruggenthaler2017b,flick2018a}. Amongst the existing first-principles methods for treating strongly coupled light-matter systems, quantum electrodynamical density-functional theory (QEDFT) has become a valuable approach for describing ground- and excited-state properties of complex matter systems coupled to a photonic environment~\cite{flick2017c,flick2019}. The Casida-like approach~\cite{casida1996} common to molecular and quantum chemistry was recently extended to a matter-photon description within the linear-response framework of QEDFT~\cite{flick2019}. The feasibility of treating the excited-state properties of a single molecule and an ensemble of molecules coupled to a realistic description of a cavity has been demonstrated~\cite{flick2020,flick2019,wang2021,sidler2021}. A different approach within QEDFT is to solve the time-dependent Kohn-Sham equations coupled to the Maxwell equations in real time~\cite{flick2018,welakuh2021t,jestaedt2020,schaefer2021}. One major advantage the real-time approach has is that it scales favorably with the system size as it involves only occupied Kohn-Sham orbitals but to obtain a converged response spectrum requires a long time-propagation, which is not favorable for larger systems. On the other hand, the Casida approach requires both occupied and unoccupied orbitals and it also scales with the number of photon modes considered.

In addition to these methods, there is another successful scheme that can combine the strengths of the previously mentioned methods known as the Sternheimer approach~\cite{sternheimer1954}. The Sternheimer approach has been used for a long time in the context of density-functional perturbation theory, for instance, for calculating phonon spectra~\cite{baroni1987a}. Recent applications of the Sternheimer equation have also been used to compute the frequency-dependent electronic response~\cite{huebener2014,huebener2014a,giustino2010,andrade2007}. The Sternheimer equation has been formulated within the framework of time-dependent density-functional theory (TDDFT) that allows to study dynamic response of much larger complex systems as it includes only occupied orbitals~\cite{andrade2007,hofmann2018}. One of a few advantages the Sternheimer approach has over real-time TDDFT is that it is formulated in the frequency space and the responses at different frequencies can be computed independently of each other allowing for the use of parallelization schemes that speed up the computation. Another advantage is that since the responses at different frequencies can be treated independently, we can compute any part of the spectrum without necessarily starting from the zero-frequency. In this work, we extend the frequency-dependent Sternheimer approach of TDDFT to the framework of QEDFT. An advantage the electron-photon Sternheimer approach has over the Casida approach is that it scales well with the system size since only occupied orbitals are treated explicitly and the arbitrarily many but finite photon modes that can be included does not add to this scaling. This approach becomes useful for investigations in polaritonic chemistry or materials in nano-plasmonic cavities that usually consider a large number of particles interacting with the electromagnetic field. We start by showing the applicability of the method in capturing not only the absorption properties of strongly coupled light-matter system but also the modified dispersion properties of the coupled system for the case of an azulene molecule. In addition, we show the spectra of the photon field which capture similar features of strong light-matter coupled systems indicating how the hybrid characteristics can be viewed from either of the subsystems at the same time highlighting the cross-talk between the subsystems. To show the efficiency of the Sternheimer approach, we study the absorption spectrum of a lithium hydride (LiH) molecule coupled to a continuum of photon modes. For the case of coupling the molecule weakly to half a million photon modes, we recover the spectrum of the free space case. By effectively enhancing the coupling of the bath modes to the molecule, we observe changes in the spectrum as the Lorentzian line shape turn into Fano resonances.

This article is structured as follows. First, we present the physical setting of a many-electron system coupled to photons in non-relativistic QED and subsequently present the linear-response setting of this framework. Secondly, we present in Sec.~\ref{sec:el-pt-sternheimer} a derivation of the frequency-dependent Sternheimer approach for electron-photon coupled systems in the linear-response regime formulated within the framework of QEDFT and discuss numerical details of the Sternheimer scheme. In the next section, we investigate the complex polarizability of a molecular system coupled to a high-Q optical cavity and highlight how the absorption and dispersion properties get modified due to strong light-matter coupling. Also, we show for the same molecular system the polaritonic features that arise in the spectra of the photon field. Furthermore, we demonstrate the efficient and low computational cost of the Sternheimer approach by coupling a LiH molecule to a (discretized) continuum of states of photon modes and show the physical effects that arises. Finally we present a conclusion and an outlook.

\section{From microscopic fields to the quantum description of light-matter interaction }
\label{sec:general-framework-1}

We are interested in the dynamics of matter interacting with the electromagnetic field within the setting of non-relativistic QED where both constituents of the coupled system are treated on an equal quantized footing. While this setting of slowly moving charged particles can be deduced from QED, concepts from classical electrodynamics are equally instructive to arrive at this description. In this regard, we lay emphasizes on the full description of the electromagnetic field that couples to a matter system. Our starting point is the inhomogeneous microscopic Maxwell equations for the transverse part of the electromagnetic field~\cite{loudon2000}:
\begin{align*}
\boldsymbol{\nabla} \times \textbf{E}(\textbf{r},t) &= - \frac{\partial}{\partial t}\textbf{B}(\textbf{r},t) \, , \\
\boldsymbol{\nabla} \times \textbf{B}(\textbf{r},t) &= \frac{1}{c^{2}} \left[\frac{\partial}{\partial t}\textbf{E}(\textbf{r},t) + \mu_{0}c^{2}\textbf{j}(\textbf{r},t)\right] \, ,
\end{align*}
where  $\textbf{E}(\textbf{r},t)$ and $\textbf{B}(\textbf{r},t)$ are the classical electric and magnetic fields, respectively. The transverse charge current $\textbf{j}(\textbf{r},t)$ represent both the free and bound current. If we consider $\textbf{j}(\textbf{r},t)$ to represent only the bound current, then it is related to the polarization $\textbf{P}(\textbf{r},t)$ of the matter as $\textbf{j}(\textbf{r},t)=\frac{\partial}{\partial t}\textbf{P}(\textbf{r},t)$. The Maxwell's equations takes into account the back-reaction of the matter on the electromagnetic field. For a quantum mechanical description, the field variables are promoted to field operators in the Heisenberg picture. In this representation, the energy of the transverse electromagnetic field can be expressed as~\cite{abedi2018}:
\begin{align}
\hat{H}_{\text{EM}} &= \frac{\epsilon_{0}}{2} \int d^{3}\textbf{r} \left[\hat{\textbf{E}}^{2}(\textbf{r}) + c^{2}\hat{\textbf{B}}^{2}(\textbf{r}) \right]  \nonumber \\
&= \frac{\epsilon_{0}}{2} \int d^{3}\textbf{r} \left[\frac{1}{\epsilon_{0}^{2}}\left(\hat{\textbf{D}}(\textbf{r}) - \hat{\textbf{P}}(\textbf{r})\right)^{2} + c^{2}\hat{\textbf{B}}^{2}(\textbf{r}) \right] , \label{eq:em-energy}
\end{align}
where we introduced the displacement field $\hat{\textbf{D}}(\textbf{r}) = \epsilon_{0}\hat{\textbf{E}}(\textbf{r}) + \hat{\textbf{P}}(\textbf{r})$. Equation (\ref{eq:em-energy}) will differ from other forms only in the choice of canonical variables and here we impose a commutation relation between $\hat{\textbf{B}}$ and $\hat{\textbf{D}}$ to be $\left[\hat{\textbf{B}}^{i}(\textbf{r}),\hat{\textbf{D}}^{j}(\textbf{r}')\right] = - i\hbar \varepsilon^{ijk}{\partial_{k}}\delta(\textbf{r} - \textbf{r}')$.
For any photonic environment of varying geometry, the fields in Eq.~(\ref{eq:em-energy}) can be expanded in the modes~\cite{abedi2018}
\begin{align}
\hat{\textbf{D}}(\textbf{r}) &=  \sum_{\alpha} \textbf{S}_{\alpha}(\textbf{r})\hat{d}_{\alpha} \, , \label{eq:displacement} \\
\hat{\textbf{B}}(\textbf{r}) &=  \sum_{\alpha} \frac{1}{\omega_{\alpha}} \hat{b}_{\alpha} \boldsymbol{\nabla} \times \textbf{S}_{\alpha}(\textbf{r}) \, , \label{eq:magnetic} \\
\hat{\textbf{P}}(\textbf{r}) &= \sum_{\alpha} \textbf{S}_{\alpha}(\textbf{r}) \left(\textbf{S}_{\alpha}(\textbf{r}_{0})\cdot\hat{\textbf{R}}\right) \, . \label{eq:polarization}
\end{align}
The expansion coefficients $\hat{d}_{\alpha}$ and $\hat{b}_{\alpha}$ are respectively the field amplitudes of the electric displacement and the magnetic field, $\textbf{S}_{\alpha}(\textbf{r})$ are the mode functions and $\hat{\textbf{R}}=\sum_{i} e \,\hat{\textbf{r}}_{i}$ is the total electronic dipole operator. In Eq.~(\ref{eq:polarization}), we employed the dipole approximation when the electromagnetic field interacts with the matter system via the electronic dipole. Making a substitution of Eqs.~(\ref{eq:displacement})-(\ref{eq:polarization}) into (\ref{eq:em-energy}) results to the following expression of the electromagnetic energy~\cite{abedi2018}:
\begin{align}
\hat{H}_{\text{EM}} &= \sum_{\alpha}\frac{1}{2}\left[\hat{p}^2_{\alpha}+\omega^2_{\alpha}\left(\hat{q}_{\alpha} \!-\! \frac{\boldsymbol{\lambda}_{\alpha}}{\omega_{\alpha}} \cdot \hat{\textbf{R}} \right)^2\right] . \label{eq:em-energy-final}
\end{align}
The displacement coordinate $\hat{q}_{\alpha}$ and momentum operator $\hat{p}_{\alpha}$ are related to the amplitudes as  $\hat{d}_{\alpha}=\sqrt{\epsilon_{0}}\omega_{\alpha}\hat{q}_{\alpha}$ and $\hat{b}_{\alpha}=\sqrt{1/\epsilon_{0}}\hat{p}_{\alpha}$ where they satisfy the commutation relation $\left[\hat{q}_{\alpha},\hat{p}_{\alpha'}\right]=i\hbar\,\delta_{\alpha,\alpha'}$. Equation (\ref{eq:em-energy-final}) tells us that what would normally be the purely photonic Hamiltonian is now a mixture of matter and photon degrees. 
The term $\boldsymbol{\lambda}_{\alpha}$ in Eq.~(\ref{eq:em-energy-final}) is the light-matter coupling strength given as
\begin{align}
\boldsymbol{\lambda}_{\alpha} = \frac{1}{\sqrt{\epsilon_{0}}} \, \textbf{S}_{\alpha}(\textbf{r}_{0}) \, , \label{coupling-strength}
\end{align}
where the mode function is evaluated at the center of charge~\cite{milonni1993}. In deriving Eq.~(\ref{eq:em-energy-final}) we have assumed a finite photonic environment with appropriate boundary conditions. For real systems we, however, have usually a continuum of modes, i.e., the cavity geometry is open to free infinite space. We can approximate this situation by extending the quantization volume of the electromagnetic field beyond the photonic environment and thus work with a discretized continuum. By making this discretization finer and finer, i.e. take the quantization volume to infinity, we can approximate the open-cavity situation arbitrarily well.
The discrete continuum description of the photon field has the advantage that it accounts for the emission or absorption of a photon in real space~\cite{flick2017} and allows for modeling an open photonic  environment~\cite{welakuh2021}. Together with the Hamiltonian representing the bound charged particles, i.e. the kinetic energy, binding and interaction potentials, equation (\ref{eq:em-energy-final})  constitutes the so-called Pauli-Fierz Hamiltonian in the length gauge~\cite{rokaj2017,schaefer2020}. In the case where we include time-dependent external perturbations, the length gauge Hamiltonian is given by
\begin{align} 
\hat{H}(t)&=\sum\limits_{i=1}^{N}\left(\frac{\hat{\textbf{p}}_{i}^{2}}{2m} + v_{\textrm{ext}}(\hat{\textbf{r}}_i,t)\right) + \sum\limits_{i>j}^{N}w(|\hat{\textbf{r}}_{i}-\hat{\textbf{r}}_{j}|)\nonumber\\
&+\sum_{\alpha=1}^{M}\frac{1}{2}\left[\hat{p}^2_{\alpha}+\omega^2_{\alpha}\left(\hat{q}_{\alpha} \!-\! \frac{\boldsymbol{\lambda}_{\alpha}}{\omega_{\alpha}} \cdot \hat{\textbf{R}} \right)^2\right] \!+\! \sum_{\alpha=1}^{M}\frac{j_{\textrm{ext}}^{(\alpha)}(t)}{\omega_\alpha}\hat{q}_\alpha.\label{el-pt-hamiltonian}
\end{align}
Here the $N$ electrons are described by the electronic coordinates $\hat{\textbf{r}}_{i}$ and the momentum operator $\hat{\textbf{p}}_{i}$, which satisfy the commutation relation $\left[\hat{\textbf{r}}_{i},\hat{\textbf{p}}_{j}\right]=i\hbar\,\delta_{ij}$. The interaction due to the longitudinal part of the photon field $w(|\hat{\textbf{r}}_{i}-\hat{\textbf{r}}_{j}|)$ can be written as a mode-expansion in Coulomb gauge which for the free-space case results to the standard Coulomb interaction
\begin{align*}
w(|\hat{\textbf{r}}-\hat{\textbf{r}}'|) = \sum_{n} \frac{e^{2}}{\textbf{k}_{n}} \frac{e^{i\textbf{k}_{n}(\hat{\textbf{r}}-\hat{\textbf{r}}') }}{\epsilon_{0}L^{3}} \xrightarrow[]{L\rightarrow\infty} \frac{e^2}{4 \pi\epsilon_0}\frac{1}{\left|\hat{\textbf{r}}-\hat{\textbf{r}}'\right|} \, ,
\end{align*}
where $\textbf{k}_{n}=2\pi\textbf{n}/L$ are the allowed wave vectors of the photon field for an arbitrarily large but finite box of length $L$~\cite{greiner1996}. For the transverse field, we consider an arbitrarily large but finite number of photon modes $M$. It is important to note that when we sample a large number of modes to describe the photon continuum, we might need to use the \textit{bare mass} of the electrons instead of the renormalized \textit{physical mass}~\cite{spohn2004,rokaj2020}. In Sec.~\ref{sec:lamb-shift} of this work, we make the common assumption that only the sampled continuum due to a cavity or photonic nanostructure is changed with respect to the free space case. The rest of the continuum of modes not affected by the cavity is subsumed in the already renormalized physical mass of the electrons. The time-dependent external potential and current in Eq.~(\ref{el-pt-hamiltonian}) can be split into 
\begin{align} 
v_{\textrm{ext}}(\textbf{r},t) &= v(\textbf{r}) + \delta v(\textbf{r},t), \nonumber \\
j_{\textrm{ext}}^{(\alpha)}(t) &= j_{\alpha} + \delta j_{\alpha}(t), \nonumber
\end{align}
where $v(\textbf{r})$ describes the attractive potentials of the nuclei and $\delta v(\textbf{r},t)$ a classical external probe field that couples to the electronic subsystem. Also, the static part $j_{\alpha}$ merely polarizes the vacuum and the time-dependent part $\delta j_{\alpha}(t)$ then generates photons in the mode $\alpha$. Either of these perturbations can be used to probe the coupled light-matter system.


\section{Linear Response formulation in the length gauge}
\label{sec:general-framework}


To characterize the properties of a system, one can investigate its response to an external perturbation. In the case of a weak external perturbation, we have access to linear response properties of the system such as its polarizability which gives access to its excitation energies and oscillator strengths. Usually this requires knowledge of the linear density response and in the case of a coupled matter-photon system, we have access to the displacement field~\cite{flick2019}. In the length gauge, the linear response of the electron density $n(\textbf{r},t)$ to the external potential $\delta v(\textbf{r},t)$ and charge current $\delta j_{\alpha}(t)$ yields the response equation~\cite{flick2019}
\begin{align}
\delta n(\textbf{r},t) &= \int dt' \int d^{3}\textbf{r}' \chi_{n}^{n}(\textbf{r},t;\textbf{r}',t') \delta v(\textbf{r}',t')  \nonumber \\
&\quad + \sum_{\alpha=1}^{M} \int dt' \chi^n_{q_\alpha}(\textbf{r},t;t') \delta j_\alpha(t') . \label{density_response}
\end{align}
Due to the coupling between light and matter, we can equally compute the linear response of the photon displacement coordinate $q_{\alpha}(t)$ due to the external potential $\delta v(\textbf{r},t)$ and current $\delta j_{\alpha}(t)$ that results to the response equation~\cite{flick2019}
\begin{align}
\delta q_\alpha(t) &= \int dt' \int d^{3}\textbf{r}' \chi^{q_\alpha}_{n}(t;\textbf{r}',t') \delta v(\textbf{r}',t') \nonumber \\ 
& \quad +  \sum_{\alpha'=1}^{M}\int dt' \chi^{q_\alpha}_{q_{\alpha'}}(t,t')\delta j_{\alpha'}(t'). \label{photon_response}
\end{align}
The response functions $\chi_{n}^{n}$, $\chi^n_{q_\alpha}$, $\chi^{q_\alpha}_{n}$ and $\chi^{q_\alpha}_{q_{\alpha'}}$ are intrinsic properties of the electron-photon coupled system which can be computed to obtain excited-state properties of the system. However, computing these response equations or the response functions directly is usually very challenging even for the electron-only system. One possible way to do this efficiently is to reformulate the response equations using the Maxwell-Kohn-Sham system of QEDFT that reproduces the same response of the density and photon coordinate~\cite{flick2015,ruggenthaler2014,flick2019}. In such a setting, the response functions can be computed approximately giving access to, for instance, excitation energies and oscillator strengths. We recently extended the Casida equation~\cite{casida1996} within the framework of QEDFT to treat electron-photon coupled systems~\cite{flick2019}. This approach computes the excitation energies and oscillator strengths of either of the coupled response functions $\left\{\chi_{n}^{n}, \chi^{q_\alpha}_{n}\right\} $ and $\left\{\chi^n_{q_\alpha}, \chi^{q_\alpha}_{q_{\alpha'}}\right\}$ by diagonalizing a pseudo-eigenvalue equation~\cite{flick2019}. The Casida approach which requires both occupied and unoccupied Kohn-Sham orbitals in addition with the sampled photon modes scales favorably for small coupled systems~\cite{sidler2021}. However, for larger electronic systems coupled to many photon modes, this leads to a rapid increase in computational effort in the Casida approach as the Casida matrix equation increases in size.

An alternative approach which rather computes the response equations instead of the response function is the frequency-dependent Sternheimer equation~\cite{sternheimer1954}. Formulated within the framework of TDDFT, this method computes the linear density response due to an external weak perturbation~\cite{andrade2007,hofmann2018,ullrich2011} as well as non-linear responses~\cite{marques2012b,andrade2007}. This approach has several advantages the main one being that it relies only on the occupied Kohn-Sham orbitals thereby restricting the computation complexity for very large systems. In the following we extend this approach to treat electron-photon coupled system within the framework of QEDFT.

\section{The Sternheimer approach for electron-photon coupled systems}
\label{sec:el-pt-sternheimer}

Practical ab-initio methods for computing optical excitation spectra are usually achieved by applying many-body methods that solve the correlated problem in an approximate way. One of such many-body methods is TDDFT which is considered a very promising methodology since it provides a good balance between accuracy and computational cost. Within the context of TDDFT there exist different formalism for computing optical excitation spectra~\cite{marques2012b}. The frequency-dependent Sternheimer method formulated within TDDFT is a perturbative approach on the Kohn-Sham orbitals that computes the density response without relying explicitly on unoccupied Kohn-Sham orbitals~\cite{andrade2007,hofmann2018}. Based on this advantage, an extension of this approach to the setting of QEDFT to treat complex atomic and molecular systems coupled to arbitrary large but finite number of photon modes is an important alternative method to existing QEDFT methods~\cite{flick2019,flick2018}. The derivation presented here is solely in the frequency space following that of Ref.~\cite{ullrich2011}. In an electron-only description, the Sternheimer approach obtains only electronic observables such as the electron density response. However, when this method is formulated within the QEDFT framework we have in addition to the density response, the response of the photon displacement coordinate (field). The mode-resolved response of the field gives access to physical processes such as the absorption or emission process. Starting with the reformulation of the density and photon displacement coordinate  responses in the QEDFT framework, the coupled response due to a weak external potential $\delta v(\textbf{r},\omega)$ are~\cite{flick2019}:
\begin{align}
\delta n(\textbf{r},\omega) &= \int d^{3}\textbf{r}'\chi_{n,s}^{n}(\textbf{r},\textbf{r}',\omega)\delta v_{\text{KS}}(\textbf{r}',\omega) \, , \label{response-by-v-omega} \\
\delta q_{\alpha}(\omega) &= \chi_{q_{\alpha,s}}^{q_{\alpha}}(\omega)\delta j_{\alpha, \text{KS}}(\omega) \, , \label{response-by-j-omega}
\end{align}
where the first-order Kohn-Sham potential and currents in Eqs.~(\ref{response-by-v-omega}) and (\ref{response-by-j-omega}) are given in terms of the interacting density and photon coordinate responses including mean-field exchange-correlation kernels:
\begin{align}
\delta v_{\text{KS}}(\textbf{r}',\omega) &= \delta v(\textbf{r}',\omega) + \int d^{3}\textbf{y} f_{\text{Mxc}}^{n}(\textbf{r}',\textbf{y},\omega) \delta n(\textbf{y},\omega) \nonumber \\
& \quad + \sum_{\alpha} f_{\text{Mxc}}^{q_{\alpha}}(\textbf{r}',\omega)\delta  q_{\alpha}(\omega) \, , \label{eff-potential-v} \\
\delta j_{\alpha, \text{KS}}(\omega)  &=  \int d^{3}\textbf{y} \, g_{\text{M}}^{n_{\alpha}}(\omega,\textbf{y}) \delta n(\textbf{y},\omega) \, . \label{eff-current-j} 
\end{align}
The mean-field exchange-correlation kernels $f_{\text{Mxc}}^{n}$ and $f_{\text{Mxc}}^{q_{\alpha}}$ are defined to be the variation of the mean-field exchange-correlation potential with respect to the density and photon coordinate, respectively while the mean-field kernel $g_{\text{M}}^{n_{\alpha}}$ is the variation of the mean-field current with respect to the electron density~\cite{flick2019}. These kernels account for the correlations in the Kohn-Sham setting of QEDFT in linear response. The non-interacting response functions of the decoupled electronic and photonic subsystems of Eq.~(\ref{response-by-v-omega}) and (\ref{response-by-j-omega}) are given explicitly as~\cite{flick2019}:
\begin{align}
\chi_{n,s}^{n}(\textbf{r},\textbf{r}',\omega) 
&= \sum_{k=1}^{N_{v}} \sum_{j=1}^{\infty} \left[ \frac{\varphi_{j}(\textbf{r})\varphi_{k}(\textbf{r}')\varphi_{k}^{*}(\textbf{r})\varphi_{j}^{*}(\textbf{r}')}{\omega - (\epsilon_{j} - \epsilon_{k}) + i\eta} \nonumber \right . \\
& \qquad\qquad \quad \left . -  \frac{\varphi_{k}(\textbf{r})\varphi_{j}(\textbf{r}') \varphi_{j}^{*}(\textbf{r})\varphi_{k}^{*}(\textbf{r}')}{\omega + (\epsilon_{j} - \epsilon_{k}) + i\eta} \right]   , \label{KS-density-response-fxn} \\
\chi_{q_{\alpha,s}}^{q_{\alpha}}(\omega) &= \frac{1}{2\omega_{\alpha}^{2}} \left(\frac{1}{\omega - \omega_{\alpha} + i\eta'} - \frac{1}{\omega + \omega_{\alpha} + i\eta'}\right) . \label{KS-photon-q-response-fxn}
\end{align}	
Here, $\epsilon_{k}$ and $\varphi_{k}(\textbf{r})$ are the ground-state energies and orbitals of the Kohn-Sham system and $\omega_{\alpha}$ is the frequency of the $\alpha$ mode. The parameters $\eta$ and $\eta'$ shifts the poles (excitation energies) of Eqs.~(\ref{KS-density-response-fxn}) and (\ref{KS-photon-q-response-fxn}) to the lower half of the complex plane and are in general not equal in both uncoupled systems.

Since the Sternheimer method is a perturbative approach on the Kohn-Sham orbitals, we start by describing the unperturbed equilibrium setting of the coupled electron-photon system as this corresponds to the zeroth-order of a perturbation expansion, for example, of the density. For this case, we start by describing the static Kohn-Sham system of ground-state QEDFT~\cite{ruggenthaler2015} where we have to solve the coupled Kohn-Sham equations
\begin{align}
\hat{h} \, \varphi_{k}(\textbf{r}) 
&= \left[\frac{\hat{\textbf{p}}^{2}}{2m} + v(\textbf{r}) + v_{\textrm{Mxc}}\left(\left[ n, q_{\alpha}\right];\textbf{r}\right)\right]\varphi_{k}(\textbf{r})  \nonumber \\
&= \epsilon_{k} \, \varphi_{k}(\textbf{r}) \, , \label{gs-electron-KS}	 \\
\omega_{\alpha}^{2} q_{\alpha} &= -\frac{1}{\omega_{\alpha}} \left(j_{\alpha} - \omega_{\alpha}^{2}  \int d^{3}\textbf{r} \, \boldsymbol{\lambda}_{\alpha}\cdot \textbf{r} \, n(\textbf{r})\right), \label{gs-photon-KS}
\end{align}
where $\hat{h}=\hat{h}\left(\left[v, n, q_{\alpha}\right]\right)$ is the ground-state Kohn-Sham Hamiltonian.  
The ground-state density can be obtained from the Kohn-Sham orbitals as $n(\textbf{r}) =\sum_{k} |\varphi_{k}(\textbf{r})|^{2}$ and the photon coordinate from Eq.~(\ref{gs-photon-KS}). The mean-field exchange-correlation potential $v_{\textrm{Mxc}}(\textbf{r})$ represents the longitudinal interactions between the electrons as well as all the transversal interactions of the electrons with the photon field.

To solve for the linear density response and photon displacement coordinate of Eq.~(\ref{response-by-v-omega}) and (\ref{response-by-j-omega}), we first start by substituting Eq.~(\ref{KS-density-response-fxn}) into the density response $n(\textbf{r},\omega)$ of Eq.~(\ref{response-by-v-omega}). The density response can now be written in a form that includes a sum over only occupied orbitals as
\begin{align}
\delta n(\textbf{r},\omega) 
&= \sum_{k=1}^{N_{v}} \left[ \varphi_{k}^{*}(\textbf{r}) \varphi_{k}^{(+)}(\textbf{r},\omega) +  \varphi_{k}(\textbf{r}) \left[\varphi_{k}^{(-)}(\textbf{r},\omega)\right]^{*}\right] , \label{density-perturbative-v}
\end{align}
where the first-order response of the Kohn-Sham orbitals $\varphi_{k}^{(\pm)}(\textbf{r},\omega)$ in Eq.~(\ref{density-perturbative-v}) are given by
\begin{align}
\varphi_{k}^{(+)}(\textbf{r},\omega) &=  \int d^{3}\textbf{r}'  \sum_{j=1}^{\infty} \frac{\varphi_{j}(\textbf{r})\varphi_{j}^{*}(\textbf{r}')\varphi_{k}(\textbf{r}')}{\omega - (\epsilon_{j} - \epsilon_{k}) + i\eta} \delta v_{\textrm{KS}}(\textbf{r}',\omega) , \label{1st-order-KS-orbital-plus}  \\
\varphi_{k}^{(-)}(\textbf{r},\omega) &= -\int d^{3}\textbf{r}'  \sum_{j=1}^{\infty} \frac{\varphi_{j}(\textbf{r})\varphi_{j}^{*}(\textbf{r}')\varphi_{k}^{*}(\textbf{r}')}{\omega + (\epsilon_{j} - \epsilon_{k}) + i\eta} \delta v_{\textrm{KS}}(\textbf{r}',\omega) . \label{1st-order-KS-orbital-minus}
\end{align}
Here, solving for the Kohn-Sham orbital responses is highly involved since we need to first determine infinitely many Kohn-Sham orbitals and evaluate an infinite sum over all these orbitals. However, this can be circumvented by acting with $\left(\omega - \hat{h} + \epsilon_{k} + i\eta\right)$ and $\left(\omega + \hat{h} - \epsilon_{k} + i\eta\right)$ on Eqs.~(\ref{1st-order-KS-orbital-plus}) and (\ref{1st-order-KS-orbital-minus}), respectively and using the static Kohn-Sham equation of Eq.~(\ref{gs-electron-KS}) in concert with the completeness of the infinite set of ground-state Kohn-Sham orbitals, i.e., $\sum_{j=1}^{\infty} \varphi_{j}(\textbf{r}')\varphi_{j}^{*}(\textbf{r}) = \delta(\textbf{r}'-\textbf{r})$. After some algebra~\cite{welakuh2021t} this leads to the frequency-dependent Sternheimer equations of the following form
\begin{align}
\left(\omega - \hat{h} + \epsilon_{k} + i\eta\right)\varphi_{k}^{(+)}(\textbf{r},\omega) &=   \delta v_{\textrm{KS}}(\textbf{r},\omega) \varphi_{k}(\textbf{r})  , \label{sternheimer-v-plus}  \\
\left(\omega + \hat{h} - \epsilon_{k} + i\eta\right)\varphi_{k}^{(-)}(\textbf{r},\omega) &= - \varphi_{k}^{*}(\textbf{r}) \delta v_{\textrm{KS}}(\textbf{r},\omega) , \label{sternheimer-v-minus}
\end{align}	
where the first-order Kohn-Sham potential $\delta v_{\textrm{KS}}(\textbf{r},\omega)$ is given by
\begin{align}
\delta v_{\textrm{KS}}(\textbf{r},\omega) &= \delta v(\textbf{r},\omega) + \int d^{3}\textbf{r}' f_{\text{Mxc}}^{n}(\textbf{r},\textbf{r}',\omega) \delta n(\textbf{r}',\omega) \nonumber \\
& \quad + \sum_{\alpha} f_{\text{Mxc}}^{q_{\alpha}}(\textbf{r},\omega)\delta  q_{\alpha}(\omega)  . \label{sternheimer-eff-potential-v}
\end{align}
The response of the photon coordinate $\delta q_{\alpha}(\omega)$ in Eq.~(\ref{response-by-v-omega}) to the external potential $\delta v(\textbf{r},\omega)$ can be expressed in the following form
\begin{align}
\delta q_{\alpha}(\omega) 
&= \delta q_{\alpha}^{(+)}(\omega) + \delta q_{\alpha}^{(-)}(\omega) , \label{photon-coordinate-v} 
\end{align} 
where we substituted Eq.~(\ref{KS-photon-q-response-fxn}) into (\ref{response-by-v-omega}). The first-order responses of the photon coordinates $\delta q_{\alpha}^{(+)}(\omega)$ and $\delta q_{\alpha}^{(-)}(\omega)$ are given explicitly as
\begin{align}
\delta q_{\alpha}^{(+)}(\omega) 
&= \frac{1}{2\omega_{\alpha}^{2}} \left(\frac{1}{\omega - \omega_{\alpha} + i\eta'} \right) \int d^{3}\textbf{r}' g_{\text{M}}^{n_{\alpha}}(\textbf{r}') \delta n(\textbf{r}',\omega), \label{sternheimer-q-v-plus} \\
\delta q_{\alpha}^{(-)}(\omega) 
&= -\frac{1}{2\omega_{\alpha}^{2}} \left( \frac{1}{\omega + \omega_{\alpha} + i\eta'}\right) \int d^{3}\textbf{r}' g_{\text{M}}^{n_{\alpha}}(\textbf{r}') \delta n(\textbf{r}',\omega). \label{sternheimer-q-v-minus}
\end{align}
To obtain the response of the density and photon coordinate of Eqs.~(\ref{density-perturbative-v}) and (\ref{photon-coordinate-v}), we have to solve Eqs.~(\ref{sternheimer-v-plus})-(\ref{sternheimer-q-v-minus}) self-consistently. The self-consistency in solving these equations becomes evident by noting that the right-hand side of the Sternheimer equations (\ref{sternheimer-v-plus}) and (\ref{sternheimer-v-minus}) depends on the solution through $\delta v_{\textrm{KS}}(\textbf{r},\omega)$ which in turn depends on $\delta n(\textbf{r},\omega)$ and $\delta q_{\alpha}(\omega)$. These two quantities depends on the first-order perturbed Kohn-Sham orbitals $\varphi_{k}^{(\pm)} (\textbf{r},\omega)$ and photon responses $\delta q_{\alpha}^{(\pm)}(\omega)$. It is important to note that the first-order response of the Kohn-Sham orbitals must satisfy the orthogonality condition with the ground-state Kohn-Sham orbitals~\cite{andrade2007,hofmann2018}:
\begin{align}
\int d^{3}\textbf{r} \, \varphi_{k}^{(*)}(\textbf{r}) \, \varphi_{k}^{(\pm)}(\textbf{r},\omega) = 0 \, .
\end{align}
From solving the self-consistent Sternheimer equations we can compute the dynamic polarizability of the coupled system which is given in terms of the variation of the density
\begin{align}
\alpha_{\mu\nu}(\omega) = \int d^{3}\textbf{r} \, \delta n_{\nu}(\textbf{r},\omega) \, \textbf{r}_{\mu} \, , \label{polarizability}
\end{align}
and is related to the photo-absorption cross-section as $\sigma(\omega)=(4\pi\omega/3c) \, \textrm{Tr} \, \bar{\alpha}_{\mu\mu}$~\cite{ullrich2011}. Since the solutions $\varphi_{k}^{\pm}(\textbf{r},\omega)$ of Eqs.~(\ref{sternheimer-q-v-minus}) and (\ref{sternheimer-q-v-plus}) are complex-valued, the density response of Eq.~(\ref{density-perturbative-v}) become complex as well. This gives rise to the polarizability $\alpha_{\mu\nu}(\omega)$ having real and imaginary parts. The imaginary part of polarizability describes the absorption of radiation, and the real part defines the refraction properties of the matter system due to a perturbation from an external  electromagnetic field~\cite{boyd1992}.

In the decoupling limit of light and matter when $|\boldsymbol{\lambda}_{\alpha}|\rightarrow0$, the Sternheimer equations (\ref{sternheimer-v-plus}) and (\ref{sternheimer-v-minus}) still retain the same form, however, the potential $\delta v_{\textrm{KS}}(\textbf{r},\omega)$ simplifies to that of an electron-only interacting system as $f_{\text{Mxc}}^{n}\rightarrow f_{\text{Hxc}}^{n}$ and $f_{\text{Mxc}}^{q_{\alpha}}\rightarrow0$. Also, the ground-state Kohn-Sham Hamiltonian in Eqs.~(\ref{sternheimer-v-plus}) and (\ref{sternheimer-v-minus}) reduces to $\hat{h}=\hat{h}([v,n])$ as $v_{\textrm{Mxc}}([n, q_{\alpha}];\textbf{r})\rightarrow v_{\textrm{Hxc}}([n];\textbf{r})$, thus, decoupling the photon contribution of Eq.~(\ref{gs-photon-KS}). The derivation of the Sternheimer scheme for the electron density and photon displacement coordinate responses in the QEDFT framework due to a weak external charge current $\delta j_{\alpha}(\omega)$ follows the same steps as above~\cite{welakuh2021t}.

Details about the numerical treatment of Eqs.~(\ref{sternheimer-v-plus}) and (\ref{sternheimer-v-minus}) have been discussed in the TDDFT framework of the frequency-dependent Sternheimer method~\cite{andrade2007,hofmann2018}. Therefore, we only summarize features in the numerical application of these equations. First, the positive infinitesimal parameter $\eta$ is required for numerical stability for the solution of the Sternheimer equations close to the resonance frequencies as it removes the divergences. It is also necessary to obtain the imaginary part of the polarizability. In addition, this parameter accounts for the artificial linewidth that represents the finite lifetimes of the excitations. Our extension of the Sternheimer method to treat electron-photon coupled systems introduced the small positive infinitesimal parameter $\eta'$ that enters the self-consistent Sternheimer equations as in Eqs.~(\ref{sternheimer-q-v-plus}) and (\ref{sternheimer-q-v-minus}). This parameter is necessary to ensure that the poles at $\omega_{\alpha}$ are finite. In our simulations we found that $\hbar\eta'=0.001$~eV is the ideal value to obtain converged results and we used $\hbar\eta=0.1$~eV. 

For the electron-photon Casida approach, the resulting dimension of the coupled but truncated matrix is $\left((N_{v}*N_{c}+M)\times(N_{v}*N_{c}+M)\right)$ where $N_{v}$ and $N_{c}$ denote the number of occupied and unoccupied Kohn-Sham orbitals, respectively~\cite{flick2019} and $M$ describes the number of photon modes. The dimensionality of the matrix increases with $N_{c}$ and $M$-photon modes. We have been so far able to treat a finite matter system coupled to $150,000$ modes 
with an efficient massive parallel implementation of the Casida equation~\cite{flick2019,welakuh2021t}. In terms of scaling with system size, the electron-photon Sternheimer approach is better when compared to the Casida approach since it still scales the same as the electron-only Sternheimer case~\cite{andrade2007,hofmann2018,huebener2014}. This is evident since we can substitute Eqs.~(\ref{photon-coordinate-v})-(\ref{sternheimer-q-v-minus}) into (\ref{sternheimer-eff-potential-v}) such that the complexity rests in solving the Sternheimer equations (\ref{sternheimer-v-plus}) and (\ref{sternheimer-v-minus}). We have implemented the linear-response frequency-dependent Sternheimer equations~(\ref{density-perturbative-v}) and (\ref{sternheimer-v-plus})-(\ref{sternheimer-q-v-minus}) into the real-space code OCTOPUS~\cite{marques2003,andrade2007}.

\section{Applications of the Frequency-dependent Sternheimer approach}
\label{sec:application-azulene}

\begin{figure}[t] 
\centerline{\includegraphics[width=0.4\textwidth]{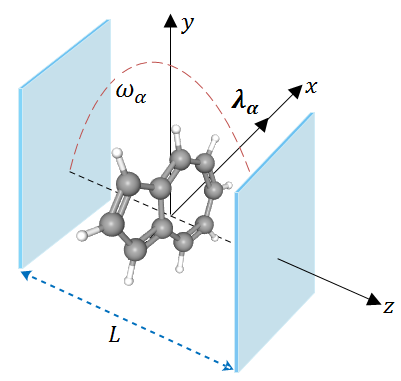}}
\caption{Schematic setup of an azulene molecule confined within a high-Q optical cavity. The cavity field is polarized along the $x$-axis with mode coupling $\boldsymbol{\lambda}_{\alpha}$ and the photon propagation vector is along the cavity axis of length $L$ in the $z$-direction. The frequency of the photon mode is $\omega_{\alpha}$.}
\label{fig:azulene-schematic}
\end{figure}

\begin{figure}[bth]
\centerline{\includegraphics[width=0.5\textwidth]{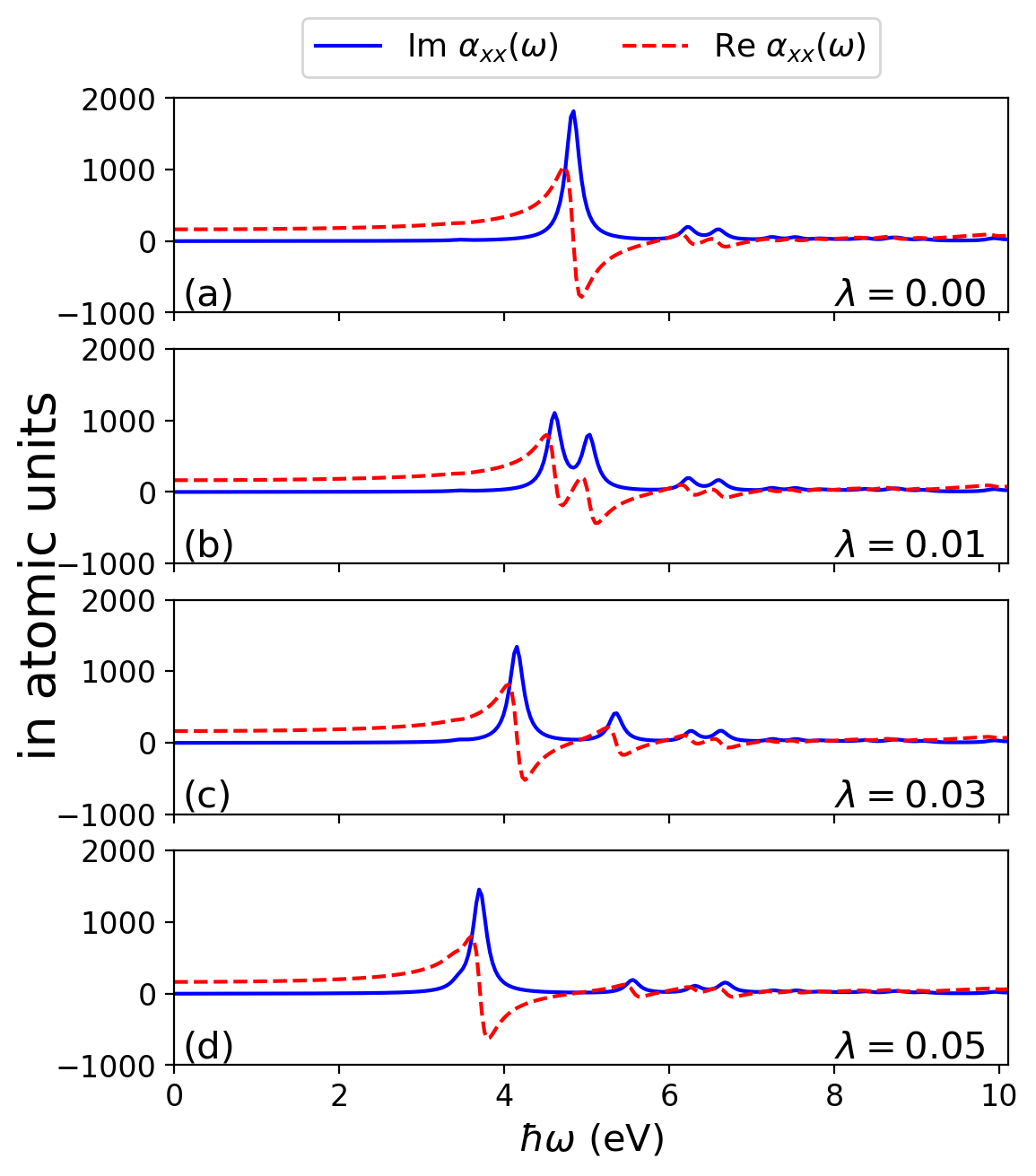}}
\caption{Spectrum of an azulene molecule in free space (i.e. $\lambda=0$) and coupled to a high-Q optical cavity (i.e. $\lambda>0$) showing the line shapes characteristic of the real and imaginary parts of the polarizability near the $\pi-\pi^{*}$ resonance at $4.825$~eV. Panel (a) shows the region near the resonance where the Re~$\left\{\alpha_{xx}(\omega)\right\}$ is asymmetric about the resonance while the Im~$\left\{\alpha_{xx}(\omega)\right\}$ is symmetric about the resonance. Coupling the cavity mode resonantly to the $\pi-\pi^{*}$ transition and increasing the coupling strength continuously as in (b) to (d) result to a Rabi splitting into lower and upper polariton branches which each have an asymmetric line shape for the different Re~$\left\{\alpha_{xx}(\omega)\right\}$.}
\label{fig:azulene-cross-section}
\end{figure}

\begin{figure}[bth]
\centerline{\includegraphics[width=0.5\textwidth]{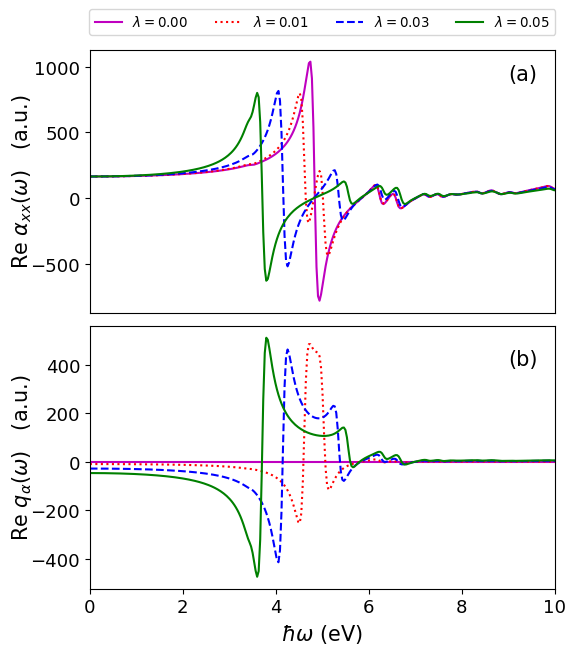}}
\caption{(a) The real part of the polarizability of azulene showing the change in the anomalous dispersion in free space ($\lambda_{\alpha}=0$) and when coupled to a cavity mode ($\lambda_{\alpha}>0$). (b) The analogous anomalous dispersion in the photon spectrum occurs only when both subsystems are coupled. Panels (a) and (b) show clearly how this feature can be controlled by coupling to a cavity mode.}
\label{fig:anomalous-dospersion}
\end{figure}

\begin{figure}[bth]
\centerline{\includegraphics[width=0.5\textwidth]{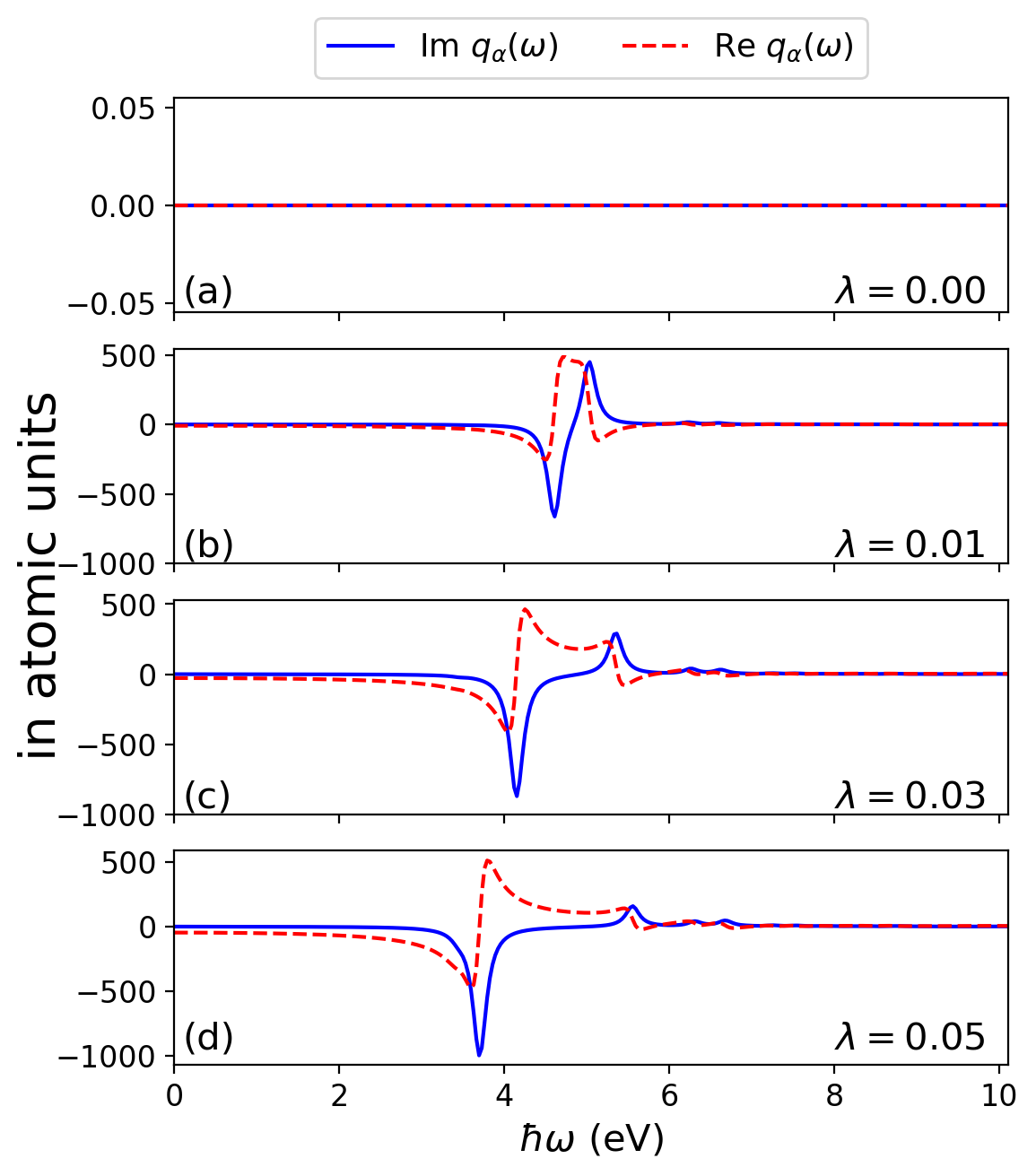}}
\caption{Spectrum of the photon displacement coordinate of an azulene molecule in free space (i.e. $\lambda=0$) and coupled to a high-Q optical cavity (i.e. $\lambda>0$). Panel (a) has no response as the photons are decoupled. Coupling the cavity mode resonantly to the $\pi-\pi^{*}$ transition and increasing the coupling strength lead to a splitting into lower and upper polaritonic branches in the photonic spectrum as shown in (b) to (d) for Im~$\left\{q_{\alpha}(\omega)\right\}$. The Re~$\left\{q_{\alpha}(\omega)\right\}$ for these cases show an antisymmetric lineshape opposite to Re~$\left\{\alpha_{xx}(\omega)\right\}$ in Fig.~\ref{fig:azulene-cross-section}.}
\label{fig:azulene-photon-coordinate}
\end{figure}

In this section, we now apply the introduced electron-photon frequency-dependent Sternheimer approach for studying excited-state properties of molecular systems coupled to a photon mode or a continuum of modes. This approach has been validated by comparing the optical absorption spectrum of a single benzene ring coupled to photons to that obtained using the electron-photon Casida and time-propagation methods of QEDFT~\cite{flick2018,welakuh2021t}. This makes the frequency-dependent Sternheimer method of QEDFT a valid alternative for studying excited states properties of strongly coupled light-matter systems.

In the following, we first investigate the physical cavity QED setup in which a single molecule is strongly coupled to a photon mode of a high-Q cavity where we expect to capture the hallmark of strong light-matter coupling (Rabi splitting). In the next setup, we include a large but finite number of photon modes that simulates the electromagnetic vacuum and investigate situations where a molecular system couples weakly and strongly to the continuum.

\subsection{Single molecule strong coupling}
\label{sec:hybrid-properties}

The first example studies intrinsic properties of a strongly coupled light-matter system that is commonly not considered, for instance, the real part of the polarizability (in Fig.~(\ref{fig:azulene-cross-section})) and the photon displacement field (in Eq.~(\ref{fig:azulene-photon-coordinate})). These quantities are particularly interesting as they give insight into the dispersive properties of the coupled system (for the real part of the polarizability) and how energy is exchanged between the electron-photon system (for the photon displacement field).

The molecular system considered here is an azulene ($\text{C}_{10}\text{H}_{8} $) molecule which is a bicyclic, nonbenzenoid aromatic hydrocarbon studied in Ref.~\cite{flick2017c}. We describe in detail how we compute the electronic structure of azulene in App.~\ref{app:azulene-system}. Before looking at how these observables get modified due to strong light-matter coupling, we will first present the absorption spectra (obtained from the imaginary part of the polarizability) of the molecular system strongly coupled to photons that captures the Rabi splitting between polaritonic peaks~\cite{flick2019,chikkaraddy2016}.

To study the spectral properties of the coupled system we now confine the azulene molecule inside an optical high-Q cavity that couples to a photon mode with increased strength. The cavity field is polarized along the $x$-direction with a coupling strength $\boldsymbol{\lambda}_{\alpha}$ as shown in Fig.~(\ref{fig:azulene-schematic}). The optical absorption spectra of the azulene molecule has been computed with TDDFT which captures the $\pi-\pi^{*}$ transition occurring at around $4.825$ eV~\cite{malloci2007,malloci2005}. In Fig.~(\ref{fig:azulene-cross-section} a.), we show the $x$-component of the polarizability of the uncoupled azulene molecule. The imaginary part of the polarizability captures a sharp peak occurring at $4.825$~eV due to the $\pi-\pi^{*}$ excitation. Based on the Kramers-Kronig relations, an absorption usually occurs simultaneously with an anomalous dispersion~\cite{boyd1992}. The anomalous dispersion describes a sudden change in the material's dispersion spectrum in the vicinity of a resonant absorption. We also find in real part of the polarizability an anomalous dispersion around the $\pi-\pi^{*}$ excitation which shows how its dispersive properties decreases when the excitation energy increases. This is characterized by the asymmetric line shape about this resonance while the imaginary part is symmetric as usually observed~\cite{bonin1997}. We now couple a single cavity mode in resonance to the $\pi-\pi^{*}$ excitation and tune the coupling $\lambda=|\boldsymbol{\lambda}|$ as in Fig.~(\ref{fig:azulene-cross-section} b-d). For the Im~$\left\{\alpha_{xx}(\omega)\right\}$, an increasing coupling strength results to an increased Rabi splitting of the $\pi-\pi^{*}$ peak into lower and upper polaritonic branches where the lower branch has more intensity, compared to the upper polaritonic peak as measured in experiments~\cite{chikkaraddy2016} and not captured by common phenomenological models such as the Jaynes-Cumming model~\cite{wang2021}. This splitting which is a characteristic of strong light-matter coupling shows how excited-state properties of matter get modified when strongly coupled to a cavity mode. For the Re~$\left\{\alpha_{xx}(\omega)\right\}$, we find for each of the lower and upper polariton peaks for different $\lambda$, asymmetric line shapes about their respective excitation energies indicating anomalous dispersion usually occurs simultaneously with absorption even for strongly coupled systems. In addition, the anomalous dispersion can be controlled for strongly coupled systems by varying the coupling strength. This is clearly shown in Fig.~(\ref{fig:anomalous-dospersion} a) where the anomalous dispersion (in particular for the lower polariton) is smaller for the coupled case when compared to the uncoupled result. The emergence of polaritonic features in the Re~$\left\{\alpha_{xx}(\omega)\right\}$ highlights that the dispersion properties of the matter system becomes modified due to strong light-matter coupling. The modification of dispersion properties for strongly coupled light-matter systems has potential in controlling optical dipole traps. This can be made clear by considering the interaction potential of the induced dipole moment normally expressed as $U = - \frac{1}{2\varepsilon_{0}c} \text{Re} \left\{ \alpha_{xx}(\omega) \right\} I$, where $I$ is the field intensity~\cite{grimm2000}. The standard approach for realizing optical dipole traps is by laser detuning from a specific resonance of the bare matter system, for instance, laser detuning from an atomic resonance such that the dipole potential minima occur at regions with maximum intensity for red-detuned traps~\cite{grimm2000}. For polaritonic resonances that emerge in strongly coupled light-matter systems, the optical dipole traps that can be realized by detuning the external field from these polaritonic resonances can be controlled by strongly coupling to the photon field. This is evident in  Fig.~(\ref{fig:azulene-cross-section}) where the Re~$\left\{\alpha_{xx}(\omega)\right\}$ is modified under strong coupling and highlights a new perspective with potential applications in engineering optical dipole traps for neutral atoms or molecules.

Next, we study the spectral properties of the photon field when we probe the matter subsystem. This observable $\delta q_{\alpha}(\omega)$ is now accessible since we treat the photon field as a dynamical part of the coupled light-matter system. We note that the displacement field in this case represents a mixed (matter and photon) spectroscopic observable since its response function $\chi_{n}^{q_{\alpha}}$ is a commutator between photonic and electronic quantities~\cite{flick2019}. The observable $\delta q_{\alpha}(\omega)$ indicates how the photon field reacts in a standard absorption or emission measurement when the system is probed by an external field represented by the potential $\delta v(\textbf{r},\omega)$. In Fig.~(\ref{fig:azulene-photon-coordinate}), we show the spectrum of the photon displacement coordinate in free space (when $\lambda=0$) and coupled to a cavity mode (when $\lambda>0$). As expected the free space case has no response since light and matter decouple and we have access only to the observables in Fig.~(\ref{fig:azulene-cross-section} a). However, coupling to the photon mode and increasing the coupling strength $\lambda>0$ we observe in the imaginary part of $\delta q_{\alpha}(\omega)$ a Rabi splitting into lower and upper polaritons peaks. The polaritonic peaks are asymmetric about the $\pi-\pi^{*}$ excitation energy to which the mode was initially coupled to and the lower polariton peaks are negative with more intensity compared to the upper polarition. Physically, this result highlights that excitations due to an external perturbation from $\delta v(\textbf{r},\omega)$ can be exchanged between the coupled subsystems and the hybrid light-matter features occur not only in the matter subsystem but also in the photon subsystem due to the self-consistent interaction. For the Re~$\left\{q_{\alpha}(\omega)\right\}$, we also find for each of the lower and upper polariton branches an asymmetric line shape about the energies of the respective polariton peaks with varying strengths for different $\lambda$. In analogy to the Re~$\left\{\alpha_{xx}(\omega)\right\}$ where the anomalous dispersion get modified due to strong light-matter coupling, the same holds true for the anomalous region in the spectrum of Re~$\left\{q_{\alpha}(\omega)\right\}$ as shown in Fig.~(\ref{fig:anomalous-dospersion} b). Due to the self-consistent back-reaction between subsystems, we expect that the Re~$\left\{q_{\alpha}(\omega)\right\}$ can be made to influence the optical dipole potential thereby controlling how the matter subsystem is trapped in the field. It is important to note that for the responses of the subsystems, the excitation energies of the strongly coupled system are the same but with differing oscillator strengths (see App.~\ref{app:azulene-system}). The results presented here demonstrates that the electron-photon Sternheimer formalism is able to describe excited-state properties of strong light-matter coupled system.

\subsection{Changes in the matter spectral features}
\label{sec:lamb-shift}

In this section, we consider the case where a molecular system is coupled explicitly to a wide range of photon modes and show how spectral features of the system change when we effectively increase its coupling to the continuum of the electromagnetic field. This computation will at the same time show the advantages the Sternheimer approach has over the Casida approach in terms of scaling with the number of photon modes.

\begin{figure}[ht] 
\centerline{\includegraphics[width=0.5\textwidth]{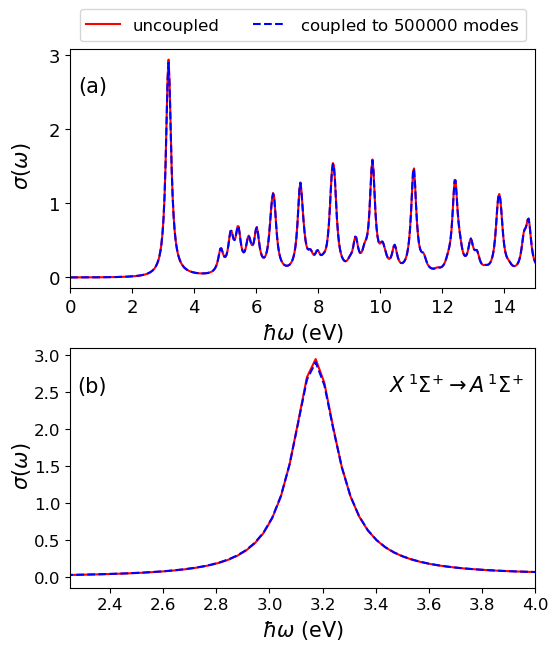}}
\caption{(a) Photo-absorption cross-section of a LiH molecule coupled to 500,000 photon modes (blue dashed) in a quasi one-dimensional cavity and its comparison to the uncoupled case (red solid). (b) Zoom-in view to the $X^{\, 1}\Sigma^{+}\rightarrow A^{\, 1}\Sigma^{+}$ transition around $3.2$~eV where we observe a slight deviation in the peak amplitude between the uncoupled and the case coupled to the continuum.}
\label{fig:lih-lifetime}
\end{figure}

\begin{figure}[ht] 
\centerline{\includegraphics[width=0.5\textwidth]{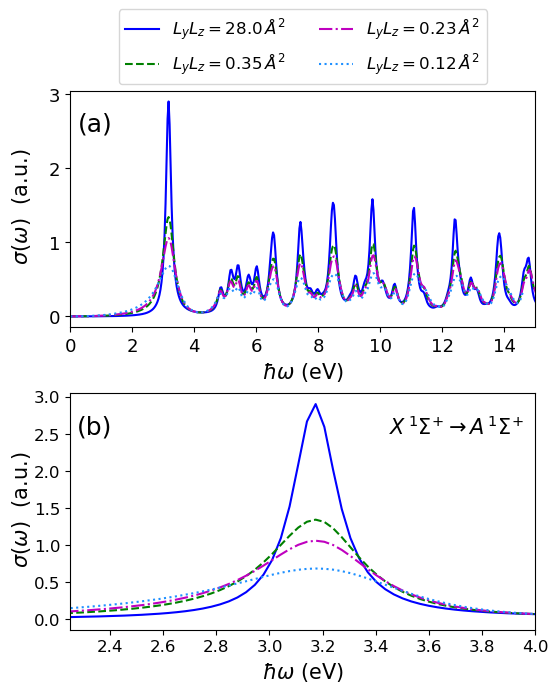}}
\caption{(a) Photo-absorption spectrum of a LiH molecule coupled to a continuum of photon modes where the coupling is effectively enhanced by changing the cavity volume via the area $L_{y}L_{z}$. The Lorentzian line shapes turn into Fano line shapes for increasing effective coupling strength. (b) Zoom-in view to the $X^{\, 1}\Sigma^{+}\rightarrow A^{\, 1}\Sigma^{+}$ transition around $3.2$~eV where we observe clearly the asymmetry of the Fano resonances when compared to Lorentzian  lineshape (blue solid line).}
\label{fig:lih-spectral}
\end{figure}

We now consider as matter system a lithium hydride (LiH) molecule coupled to a wide range of photon modes that sample densely the electromagnetic vacuum. Since the Sternheimer approach for electron-only system is known to scale favorably with the system size~\cite{andrade2007,hofmann2018}, the focus here will be to demonstrate that the photon modes do not add to this scaling. Here we sample modes of a quasi one-dimensional mode space by employing the coupling $\boldsymbol\lambda_\alpha = \sqrt{\frac{2}{\epsilon_0 L_xL_y L_z}} \text{sin}( \omega_\alpha/c  \, x_0)\textbf{e}_x$, where $x_0 = L_x/2$ is the position of the molecule in $x$-direction and $\omega_\alpha = \alpha c\pi /L_x$ are the frequencies of the modes~\cite{flick2019}. The volume $V=L_{x}L_{y}L_{z}$ with $L_{x}=3250$ $\mu$m, $L_{y}=10.58$ \AA ~and $L_{z} =2.65$ \AA ~are chosen such that the sampled modes couple weakly to the molecular system and we assume a constant mode function in the $y$ and $z$-directions.

In this first example, we couple the molecule to 500,000 photon modes of a one-dimensional mode space with an energy cut-off of 190.74~eV and a spacing between modes of 0.38~meV. Sampling the continuum of modes serve to constitute the line width of the excitations and also represent dissipation channels in the coupled system~\cite{welakuh2021,flick2019}. The one-dimensional sampling of mode frequencies that couple weakly to the matter subsystem will not capture the actual three-dimensional lifetimes. In the matter-only (uncoupled) case, we use a broadening $\hbar\eta=0.1$~eV (as in Sec.~\ref{sec:hybrid-properties}) to account for the finite lifetime of the excited states. When the molecule is coupled weakly to the photon continuum, we obtain the same spectral broadening as the uncoupled case. The results of this calculation is shown in Fig.~(\ref{fig:lih-lifetime}) where we compare the photo-absorption cross-section of the uncoupled LiH molecule and the case when it is weakly coupled to 500,000 photon modes. We find that the two results are qualitatively the same which is evident for the lowest electronic transition $X^{\, 1}\Sigma^{+}\rightarrow A^{\, 1}\Sigma^{+}$ around $3.2$~eV that corresponds to an electronic transition from the bonding to the antibonding $\sigma$-orbital~\cite{villaume2010,stwalley1993}. This result shows that the weak coupling of the molecule to the continuum of modes reproduces the results of the matter-only case. We note that obtaining this result using the electron-photon Casida approach will increase the computational cost drastically even for the case of coupling to 100,000 photon modes. Computationally, this result demonstrates that the electron-photon Sternheimer method scales favorably not only with system size but also with the number of photon modes.

Now, we effectively enhance the coupling strength $|\boldsymbol{\lambda}|$ by reducing the cavity volume along the $y$- and $z$-direction. The results are shown in Fig.~\ref{fig:lih-spectral} where the blue line is the result shown in Fig.~\ref{fig:lih-lifetime} that has a Lorentzian profile. We find that upon reducing the area $L_{y}L_{z}$, this effectively enhances the coupling to the photon continuum such that the symmetric Lorentz line shapes turn into asymmetric Fano line shapes. Fano resonances occur due to the interference of discrete quantum states with a continuum of states~\cite{ott2013,limonov2017}. The asymmetry is characterized as a ratio of the transition amplitude to a given discrete state and that of a transition to a continuum state~\cite{miroshnichenko2010}. As this ratio becomes finite due to strong coupling to the continuum, this indicates the onset of a competition between constructive and destructive interference that gives rise to the asymmetric line shape~\cite{welakuh2022a}. Also, the broadening of the spectra (see Fig.~(\ref{fig:lih-lifetime} b)) and decrease in amplitude is a consequence of the interference~\cite{welakuh2022a}. These results show the changes in the spectral features of excited-states of a matter system strongly coupled to the electromagnetic continuum. Thus, the electron-photon Sternheimer approach is a valid alternative method for studying excited-state properties of real systems strongly interacting with the quantized electromagnetic field.

\section{Conclusion and outlook}
\label{sec:conclusion-outlook}

In this work we presented a linear-response method that solves the response equations of non-relativistic QED in the length gauge setting. The approach is based on the Sternheimer equation formulated within the framework of QEDFT that is capable of computing excited-state properties of strongly coupled light-matter systems. This approach serves as an alternative linear-response method for studying response properties of large systems coupled to the quantized electromagnetic field since it scales favorably with the system size as it utilizes only the occupied Kohn-Sham orbitals and it also scales favorably with the number of photon modes. Using the Sternheimer approach we computed different observables of strongly coupled systems. These observables showed how both the dispersion and absorption properties of the matter system changes with potential applications in modifying and controlling optical dipole traps. Also, we showed examples where we lift the restriction to one cavity mode in the dipole approximation and sampled densely the electromagnetic continuum. In one case we showed that when a LiH molecule is weakly coupled to the photon continuum, we reproduce the free space absorption spectrum of the molecule. When the coupling strength between the light and matter is effectively enhanced, we find changes in the absorption spectrum as symmetric Lorentzian line shapes turn into asymmetric Fano line shapes. 

The electron-photon Sternheimer method presented here is a suitable approach for studying excited-state properties of large systems coupled to a single mode or to the electromagnetic continuum. In the fast-growing field of polaritonic chemistry where there is an ongoing debate about the mesoscopic scale of quantum-collectively of coupled molecules~\cite{sidler2021,sidler2021a}, ab-initio methods such as the electron-photon Sternheimer method become desirable to capture intricate details of the complex interactions between the coupled subsystems. Another important property of the Sternheimer approach is that it can be generalized to higher orders to obtain higher order polarizabilities by solving a hierarchy of Sternheimer equations~\cite{andrade2007}. For the coupled electron-photon system, this will give access to higher order polarizabilities with signatures of strong light-matter coupling. 


\section{Acknowledgments}

We acknowledge financial support from the European Research Council (ERC-2015-AdG-694097) and the SFB925 "Light induced dynamics and control of correlated quantum systems". This work was supported by the Excellence Cluster "CUI: Advanced Imaging of Matter" of the Deutsche Forschungsgemeinschaft (DFG), EXC 2056, project ID 390715994.  The Flatiron Institute is a division of the Simons Foundation.

\appendix

\section{Photo-absorption cross-section of the azulene molecule}%
\label{app:azulene-system}

\begin{figure}[bth]
\centerline{\includegraphics[width=0.5\textwidth]{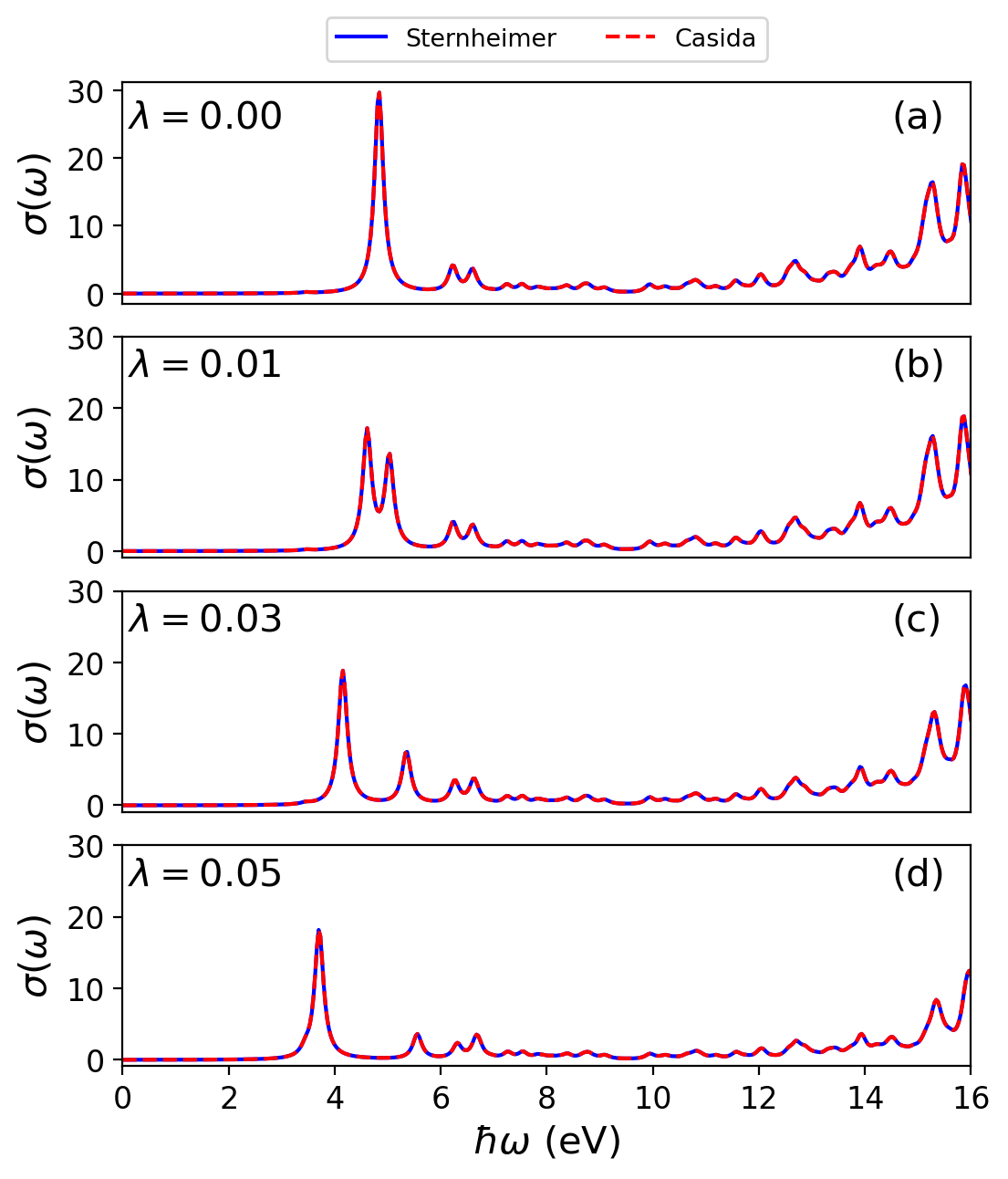}}
\caption{Comparison of the photo-absorption cross-section of an azulene molecule using the electron-photon Sternheimer and Casida methods. (a) Absorption cross-section in free space (i.e. $\lambda_{\alpha}=0$). Coupling the cavity mode resonantly to the $\pi-\pi^{*}$ transition and increasing the coupling strength continuously as in (b) to (d) leads to a Rabi splitting into lower and upper polariton branches. Both approaches are in good agreement for the computed spectra.}
\label{fig:azulene-cross-section-2}
\end{figure}

\begin{figure}[bth]
\centerline{\includegraphics[width=0.5\textwidth]{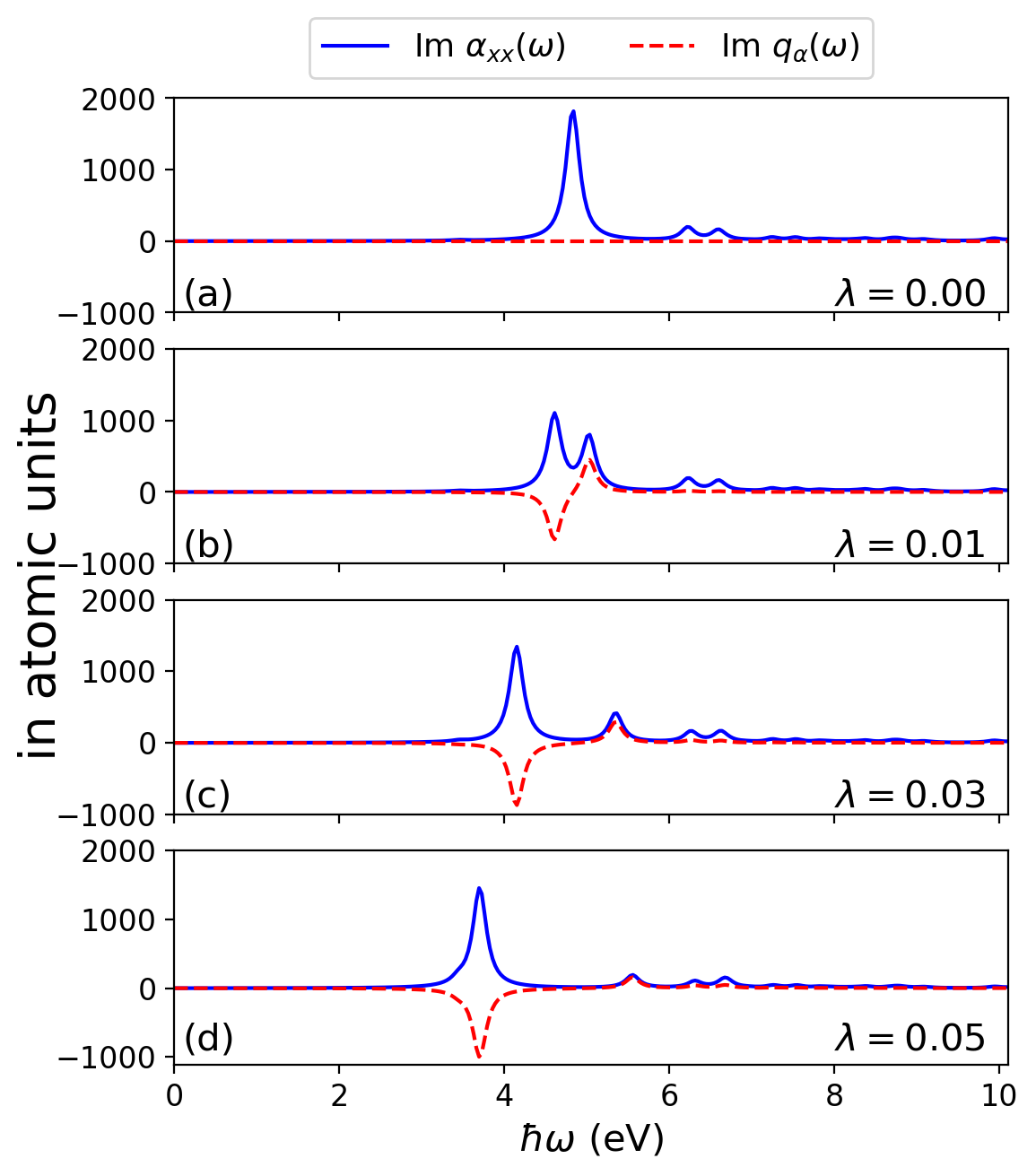}}
\caption{The spectrum of the imaginary parts of the polarizability and photon displacement coordinate of an azulene molecule in free space and in an optical cavity. (a) The free space case where light decouples from matter and there is no response from the photon subsystem. Panels (b)-(d) shows the light-matter coupled case where both observables are shown to have the same excitation energies but different oscillator strengths.}
\label{fig:azulene-polar-photon-q}
\end{figure}

To describe the azulene molecule studied in Sec.~\ref{sec:application-azulene}, we compute the electronic structure of the azulene molecule using the same setup as that in Ref.~\cite{flick2019} for the benzene molecule. That is, a cylindrical real space grid of $8$ \AA~in length with the radius of 6 \AA~in the $x$-$y$ plane and a spacing $\Delta x = \Delta y = \Delta z = 0.22$ \AA~is used. We treat the core electrons of the carbon and hydrogen atoms using the local-density approximation Troullier-Martins pseudopotentials~\cite{troullier1993}. Since the Sternheimer method considers only occupied orbitals, the $48$ valence electrons that amounts to $24$ doubly occupied orbitals are used in the computation. When we use the Casida approach, we include 500 unoccupied states which amounts to $N_{v}*N_{c}=24*500=12000$ pairs of occupied-unoccupied states. To account for the electron-electron and electron-photon interaction in the linear-response QEDFT framework~\cite{flick2019}, we apply the adiabatic LDA (ALDA) to the Hartree exchange-correlation kernel (i.e. $f^{n}_{\text{Hxc}} \rightarrow f^{n}_\text{Hxc,ALDA}$) and the photon random-phase approximation to the photon exchange-correlation kernels (i.e. $f^{n}_{\text{pxc}}, f^{q_{\alpha}}_{\text{pxc}} \rightarrow f^{n}_\text{p}, f^{q_{\alpha}}_\text{p}$), respectively. We compute the photo-absorption cross-section of the molecule in free space and inside the high-Q optical cavity and the results are shown in Fig.~(\ref{fig:azulene-cross-section-2}. This result is related to the imaginary part of Fig.~(\ref{fig:azulene-cross-section}) and it shows more excitations at higher energies.

In the linear-response setting of non-relativistic QED in the length gauge~\cite{flick2019}, the response quantities that arise have the same excitation energies but different oscillator strengths as shown in Fig.~(\ref{fig:azulene-polar-photon-q}) for two observables of interest. This is because their respective response functions have the same excitation energies and different oscillator strengths~\cite{flick2019}.

\vspace{10em}

\bibliography{01_light_matter_coupling} 
\newpage

\end{document}